\documentclass[a4paper,notitlepage,twocolumn,footnoteinbib,superscriptaddress,longbibliography,10pt]{revtex4-1}
\usepackage{amsmath,bbold,amssymb,amsthm,mathtools,color,listings,graphicx}
\usepackage[colorlinks=true,linkcolor=blue,citecolor=blue]{hyperref}

\begin{document}

\title{Multidimensional hyperspin machine}
\author{Marcello Calvanese Strinati}
\email{marcello.calvanesestrinati@gmail.com}
\affiliation{Centro Ricerche Enrico Fermi (CREF), Via Panisperna 89a, 00184 Rome, Italy}
\author{Claudio Conti}
\affiliation{Physics Department, Sapienza University of Rome, 00185 Rome, Italy}
\affiliation{Institute for Complex Systems, National Research Council (ISC-CNR), 00185 Rome, Italy}
\affiliation{Centro Ricerche Enrico Fermi (CREF), Via Panisperna 89a, 00184 Rome, Italy}
\date{\today}

\begin{abstract}
From condensed matter to quantum chromodynamics, multidimensional spins are a fundamental paradigm, with a pivotal role in combinatorial optimization and machine learning. Machines formed by coupled parametric oscillators can simulate spin models, but only for Ising or low-dimensional spins. Currently, machines implementing arbitrary dimensions remain a challenge. Here, we introduce and validate a hyperspin machine to simulate multidimensional continuous spin models. We realize high-dimensional spins by pumping groups of parametric oscillators, and study NP-hard graphs of hyperspins. The hyperspin machine can interpolate between different dimensions by tuning the coupling topology, a strategy that we call ``dimensional annealing''. When interpolating between the XY and the Ising model, the dimensional annealing impressively increases the success probability compared to conventional Ising simulators. Hyperspin machines are a new computational model for combinatorial optimization. They can be realized by off-the-shelf hardware for ultrafast, large-scale applications in classical and quantum computing, condensed-matter physics, and fundamental studies.
\end{abstract}

\maketitle

Systems of interacting spins are ubiquitous in nature. Their complex collective behavior and their equilibrium properties describe magnetism in solid-state systems~\cite{grosso2013solid}, phase transitions in spin glasses~\cite{parisi1987spinglass}, quantum chromodynamics (QCD)~\cite{PELISSETTO2002549}, and quantum and classical computation~\cite{De_las_Cuevas_2009}. Spin models are also pivotal in combinatorial optimization~\cite{10.3389/fphy.2014.00005}, with applications in machine learning~\cite{date2021qubo}, traffic and portfolio optimization~\cite{01605682.2020.1718019}, markets and finance~\cite{gilli2011numerical}, biology and life science~\cite{zhanglifecience}, artificial intelligence~\cite{ohzechiarticifialintellingence}, protein folding~\cite{winfree2002protein}, epidemic spreading~\cite{zhou2020minimization}, bioinformatic~\cite{naturecombinatorial}, and material engineering~\cite{PhysRevLett.114.105503}.

However, simulating and understanding spin systems remain a challenge, as several models are computationally (NP-)hard~\cite{Barahona_1982}. Novel algorithms and techniques are emerging, including the realization of specialized physical machines that converge to the ground state (GS) of programmable spin Hamiltonians, a major quest in the last decades~\cite{canals2019magnetic}. However, most of the work has been limited to the simulation of one-component, discrete spin systems (Ising model), and of continuous spin models with two or three components (XY or Heisenberg models, respectively). Also, spin machines, either software or hardware, suffer of limitations as heterogeneity and stiffness, which reduce the success probability in computationally NP-hard models to a narrow range of parameters. Heterogeneity refers to the fact that the spin simulator exhibits local energy minima not present in the target model. Stiffness appears in binary models that display deep local energetic states, which impede reaching the ground state during minimization or annealing. Ideally, one would use high dimensional systems to increase the symmetry in order to connect the many local minima and interpolate binary spins with continuous variables to exit the energetic traps. However, these features are not available at the moment.

\begin{figure*}
\centering
\includegraphics[width=17.0cm]{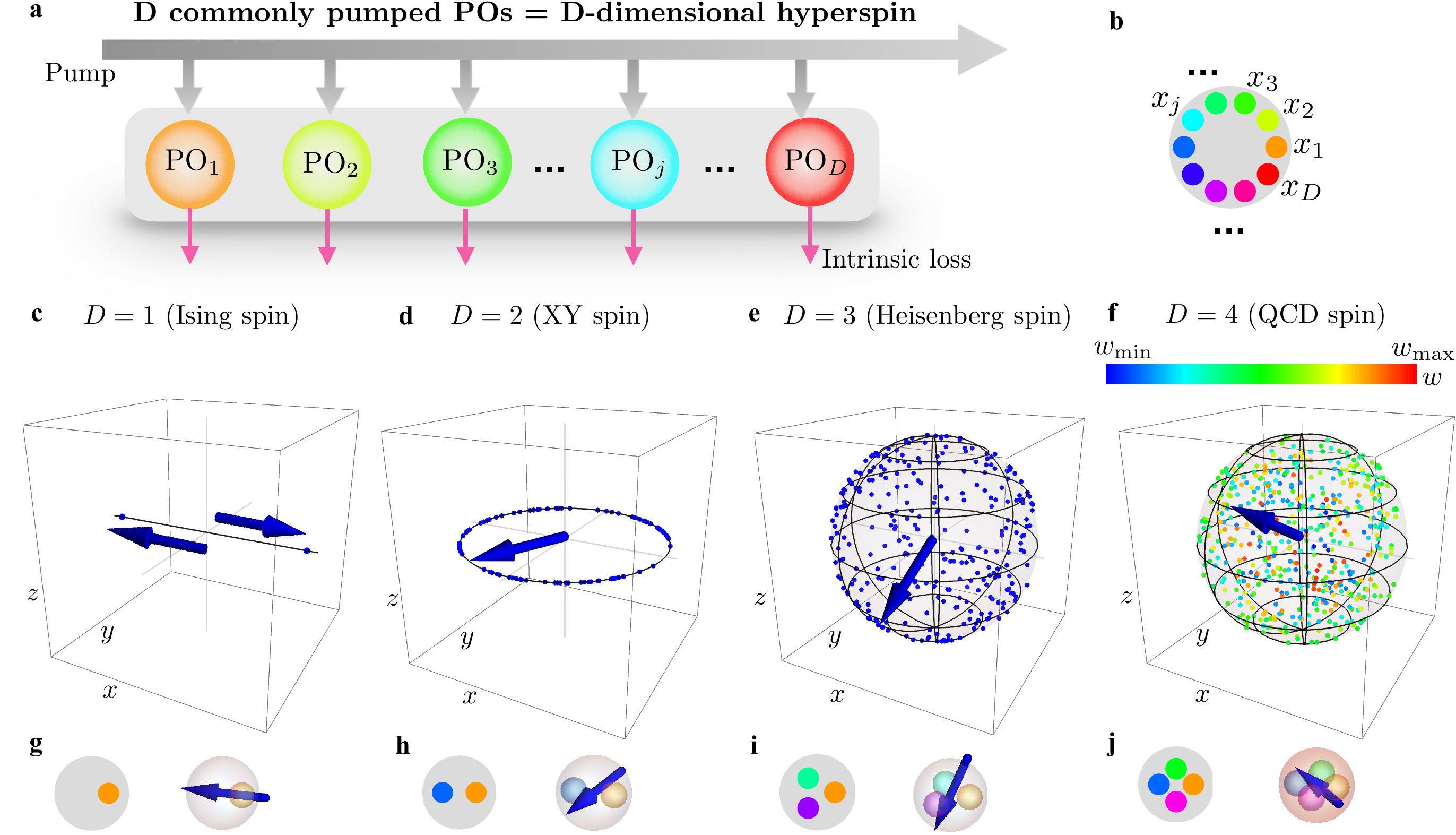}
\caption{\textbf{a}, Composite parametric oscillator (PO) as a $D$-dimensional hyperspin. The spin consists of $D$ degenerate POs (colored dots) saturating the same pump field (gray arrows and area) with equal intrinsic loss (purple arrows). \textbf{b}, Compact representation of the scheme in \textbf{a}. The colored dots denote the $D$ POs, described by dynamical variables $x_1,\ldots x_D$, and the gray circle represents the common pump. \textbf{c,d,e,f}, Fixed points of the composite PO system. The fixed points lie on the surface of a $D$-dimensional hypersphere. For $D=1$, there are two fixed points on the $x$-axis representing the two states of an Ising spin; For $D=2$, the fixed points lie on a circumference in the $xy$-plane, encoding the continuous phase of a XY spin; For $D=3$, the fixed points lie on the surface of a sphere in the $xyz$-space, encoding the two angles of an Heisenberg spin; For $D=4$, the fixed points are represented by encoding three of the four coordinates into a point within the volume of a sphere in the $xyz$-space, and the fourth coordinate $w$ is encoded as a color with extremal values $w_{\rm min}$ and $w_{\rm max}$ in the colormap. \textbf{g,h,i,j}, Left panels are the composite PO representation of the $D$-dimensional spin as in panel \textbf{b}, while right panels are a three-dimensional representation of the composite PO as a spin $\vec\sigma$ in standard hyperspherical coordinates (blue arrow). For $D=4$, borrowing the terminology from QCD, the arrow represents the ``meson'' $xyz$-component of the QCD spin, and the color assigned to the outer sphere encodes the ``scalar''  $w$-component.}
\label{fig:opomultiplet1}
\end{figure*}

Ising spin simulators include two-component Bose-Einstein condensates~\cite{Byrnes_2011,yamamotoboseeinstein2013}, superconducting circuits~\cite{Johnson2011}, digital computers~\cite{Tiunov:19,hgoto2019cim,hgoto2021cim}, electrical oscillators~\cite{chou2019}, optoelectronical oscillators~\cite{bohm2019}, and degenerate optical parametric oscillators (POs)~\cite{PhysRevA.88.063853,marandi2014cim,hamerlyfristratedchain2016,PhysRevA.96.043850,PhysRevLett.122.213902,PhysRevLett.123.083901,10.1007/978-3-030-19311-9_19,gaeta2020,Pierangeli:20,PhysRevA.104.013715} forming a coherent Ising machine (CIM). Proposed platforms to simulate classical XY models include laser networks~\cite{PhysRevResearch.2.033008,PhysRevResearch.2.043335}, non-degenerate POs~\cite{Takeda_2017}, and polariton condensates~\cite{kalininpolariton2017}. Quantum spin simulators include trapped atomic ion crystals for the quantum Ising, XY, and Heisenberg models~\cite{monroequantumsimulation2010,bollingerengineering2012,RevModPhys.93.025001,PhysRevA.104.013302}.

Programmable multicomponent spins represent a toolbox for the systematic study of nontrivial phases, symmetry breaking phenomena, as well as critical behaviours of phase transitions in condensed-matter physics~\cite{PELISSETTO2002549}. Examples include magnetic properties of three dimensional spins~\cite{PhysRevB.103.174422} and critical properties of spin glasses~\cite{Baity_Jesi_2019}. Importantly, classical multidimensional spins may simulate the behaviour of quantum many-body systems~\cite{PhysRevB.104.054415}. Four-dimensional spin models appear in QCD, describing the symmetry and critical properties of the chiral phase transition with two light-quark flavors~\cite{PhysRevD.29.338,PELISSETTO2002549,PhysRevD.85.094506,PhysRevLett.123.062002}.

In this article, we propose and validate a classical simulator of a $N$-spin system with an arbitrary number $D$ of spin components. Our proposal employs $D$ nonlinear POs to construct a multidimensional spin (or \emph{hyperspin}), achieved by driving groups of POs with a common pump field, where a single PO represents a component of the hyperspin. Different multidimensional spin Hamiltonians can be simulated by coupling POs in a hierarchical topology. The choice of POs as fundamental constituents of an hyperspin is motivated by the fact that they furnish a versatile platform to realize artificial spin devices at room temperature. POs grant an extraordinary degree of control and the prospect to realize scalable systems of coupled all-optical POs with size-independent ultra-fast equilibration times~\cite{PhysRevApplied.16.054022}. However, despite their potential use as physical hardware, we show here that even only the software implementation of an hyperspin machine enables a novel strategy for annealing that we call ``dimensional annealing'', which increases the probability to optimize hard models in a wide range of parameters.

\vspace{0.4cm}
\noindent
\textbf{Multidimensional hyperspin with POs}\\
Figure~\ref{fig:opomultiplet1}\textbf{a,b} show the construction of a $D$-dimensional spin from $D$ degenerate POs. We consider $D$ identical POs, all with frequency $\omega_0$ and loss $g$, described by classical dynamical variables $x_1,\ldots x_D$ pumped by an external drive with amplitude $h$ and frequency $2\omega_0$. The pump feeds the $D$ oscillators, with saturation value $h(1-\beta I)$, where $\beta$ is a saturation coefficient and $I$ is the total PO energy. We describe the dynamics by $D$ coupled Mathieu's equations~\cite{PhysRevLett.123.083901,PhysRevA.100.023835,Calvanese_Strinati_2020}
\begin{equation}
\ddot x_j\!+\!\omega_0^2\!\left[1\!+\!gh\!\left(1\!-\!\beta\sum_{l=1}^{D}x^2_l\right)\sin(2\omega_0t)\right]\!x_j\!+\!\omega_0g\dot x_j\!=\!0\,\,,
\label{eq:equationofmotionmultiplets1bis0}
\end{equation}
where $j=1,\ldots,D$ labels the different POs. When pumped above the threshold value $h_{\rm th}$, each PO $x_j$ responds with an oscillation at frequency locked to half the pump frequency due to period doubling instability. This oscillation is modulated by a complex amplitude $X_j$, which describes the nontrivial dynamics of the PO variable $x_j$. When an amplitude steady state exists, the fixed points $\overline{X}_j=|\overline{X}_j|e^{i\phi_j}$ encode the equilibrium values of the magnitude and phases of the PO fast oscillations, $\overline{x}_j(t)=2|\overline{X}_j|\cos(2\omega_0t+\phi_j)$.

The dynamics of the complex amplitudes $X_j$ is found from Eq.~\eqref{eq:equationofmotionmultiplets1bis0} by a multiple-scale expansion~\cite{kevorkian1996multiple,PhysRevA.100.023835}, as detailed in the supplementary information (SI). For a range of $h$ values above threshold, the dynamics amplifies the amplitude real parts, and suppresses the imaginary parts. The fixed-point values $\overline{X}_j\coloneqq\lim_{t\rightarrow\infty}X_j(t)$ are real numbers, i.e., the phase $\phi_j$ is binary (either $0$ or $\pi$). In units such that $\omega_0=1$, the time evolution of the amplitudes reads
\begin{equation}
\frac{\partial X_j}{\partial t}=\left(\frac{h}{4}-\frac{1}{2}-\frac{h\beta}{2}\sum_{l=1}^{D}X^2_l\right)X_j \,\, .
\label{eq:equationofmotionmultiplets4bisbis0}
\end{equation}
The reason why the PO system in Eq.~\eqref{eq:equationofmotionmultiplets4bisbis0} can describe a $D$-dimensional spin follows from the fixed point configuration of the amplitude dynamics, which are found as customary by equating Eq.~\eqref{eq:equationofmotionmultiplets4bisbis0} to zero. This implies $\sum_{l=1}^{D}\overline{X}^2_l=S^2$ where $S=\sqrt{\sum_{l=1}^{D}\overline{X}^2_l}=\sqrt{(1/2-1/h)/\beta}$. Thus, $\{\overline{X}_j\}_{j=1}^{D}$ are the Cartesian coordinates of a point on a $D$-dimensional hypersphere, and the corresponding unit vector is a continuous, $D$-dimensional hyperspin, i.e., $\vec\sigma=(\overline{X}_1,\ldots,\overline{X}_{D})/S$.

\begin{table}[t]
\centering
\begin{tabular}{|c|c|c|}
\hline
\hspace{0.2cm} \textbf{Dimension} \hspace{0.2cm} & \hspace{0.2cm} \textbf{OPO quadratures} \hspace{0.2cm} & \hspace{0.2cm} \textbf{Spin $S\vec{\sigma}$} \hspace{0.2cm} \\\hline
$1$ & $\overline{X}_{1}$ & Ising\\\hline
$2$ & $(\overline{X}_{1},\overline{X}_{2})$ & XY\\\hline
$3$ & $(\overline{X}_{1},\overline{X}_{2},\overline{X}_{3})$ & Heisenberg\\\hline
$4$ & $(\overline{X}_{1},\overline{X}_{2},\overline{X}_{3},\overline{X}_{4})$ & QCD\\\hline
\end{tabular}
\caption{Fixed-point PO quadratures $\overline{X}_1,\ldots,\overline{X}_D$ as a $D$-dimensional hyperspin $\vec{\sigma}\coloneqq(\overline{X}_1,\ldots,\overline{X}_D)/S$ for Ising spin ($D=1$), XY spin ($D=2$), Heisenberg spin ($D=3$), and QCD spin ($D=4$). The QCD spin is shown as a three-dimensional vector $(\overline{X}_{1},\overline{X}_{2},\overline{X}_4)$ and a scalar $\overline{X}_3$, representing the meson and scalar component, respectively (Fig.~\ref{fig:opomultiplet1}).}
\label{tab:oposandspindimension1}
\end{table}

To clarify the connection between the PO system in Eq.~\eqref{eq:equationofmotionmultiplets1bis0} and a continuous $D$-dimensional spin, we show in Fig.~\ref{fig:opomultiplet1}\textbf{c,d,e,f} the configuration of the fixed points for the specific cases $D=1,2,3,4$. We numerically integrate Eq.~\eqref{eq:equationofmotionmultiplets4bisbis0} using different random initial conditions. At the end of each integration, we obtain the real coordinates $\{\overline{X}_j\}_{j=1}^{D}$, and plot them in the $xyz$-space as blue or colored dots as follows: For $D=1$ (panel \textbf{c}), one has a single PO with two fixed points that describe the two values of an Ising spin (see Table~\ref{tab:oposandspindimension1}). In our notation, the PO fixed-point quadrature identifies the $x$-coordinate, and the $y$- and $z$- coordinates are set to zero. For $D=2$ (panel \textbf{d}), the two quadratures take any value on a circumference, and they identify the $x$- and $y$- coordinates (the $z$-coordinate is set to zero). The corresponding unit vector defines an XY spin. For $D=3$ (panel \textbf{e}), the three PO quadratures take any value on the surface of a sphere (the $z$-coordinate being identified by $\overline{X}_{3}$), and the unit vector defines an Heisenberg spin. For $D=4$ (panel \textbf{f}), a fixed point has four coordinates on the surface of a four-dimensional hypersphere. We plot the three-dimensional projected vector $(\overline{X}_{1},\overline{X}_{2},\overline{X}_4)$ within the volume of a three-dimensional sphere of radius $S$, and the extra quadrature $\overline{X}_3$ defines the fourth coordinate $w$ whose value is encoded as a color. Following the conventionally adopted terminology in QCD, we name the four-dimensional unit vector as a QCD spin, where the projected vector in the $xyz$-space is the ``meson'' and $\overline{X}_3$ is the ``scalar'' field~\cite{TETRADIS200393,SCHAEFER2005479,PhysRevD.73.074010,PhysRevD.85.094506}.

\begin{figure}
\centering
\includegraphics[width=7.5cm]{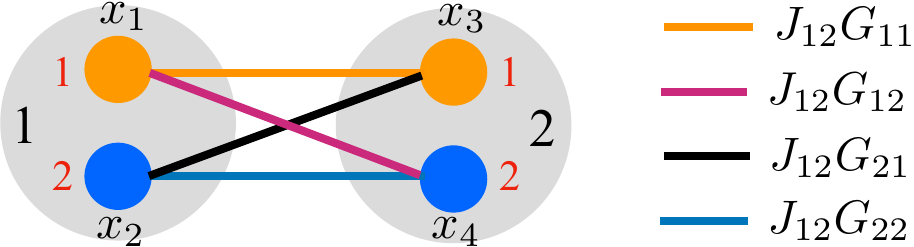}
\caption{PO connectivity as hyperspins with $N=D=2$. The POs $x_j$ with $j=1,2,3,4$ form two spins labelled by $q=1,2$ (black labels) with two components $\mu=1,2$ each (red labels), where the indexes are related as $\mu=1+(j-1){\rm mod}(D)$ and $q=1+\lfloor(j-1)/D\rfloor$. The indexes $j$ are then grouped as $\mathbb{S}_1=\{1,2\}$ and $\mathbb{S}_2=\{3,4\}$. The coupling term $C_{jl}$ between $x_j$ and $x_l$ is decomposed as $C_{13}=J_{12}G_{11}$, $C_{14}=J_{12}G_{12}$, $C_{23}=J_{12}G_{21}$, and $C_{24}=J_{12}G_{22}$ (see legend), while $C_{12}=C_{23}=0$. The amplitudes $X_1\equiv X^{(1)}_1$ and $X_2\equiv X^{(1)}_2$, and $X_3\equiv X^{(2)}_1$ and $X_4\equiv X^{(2)}_2$ form the $\mu=1,2$ components of the $q=1$ and $q=2$ hyperspins, respectively, $\vec{S}_1=(X_1,X_2)$ and $\vec S_2=(X_3,X_4)$.}
\label{fig:opomultiplet2a}
\end{figure}

\begin{figure*}
\centering
\includegraphics[width=17.9cm]{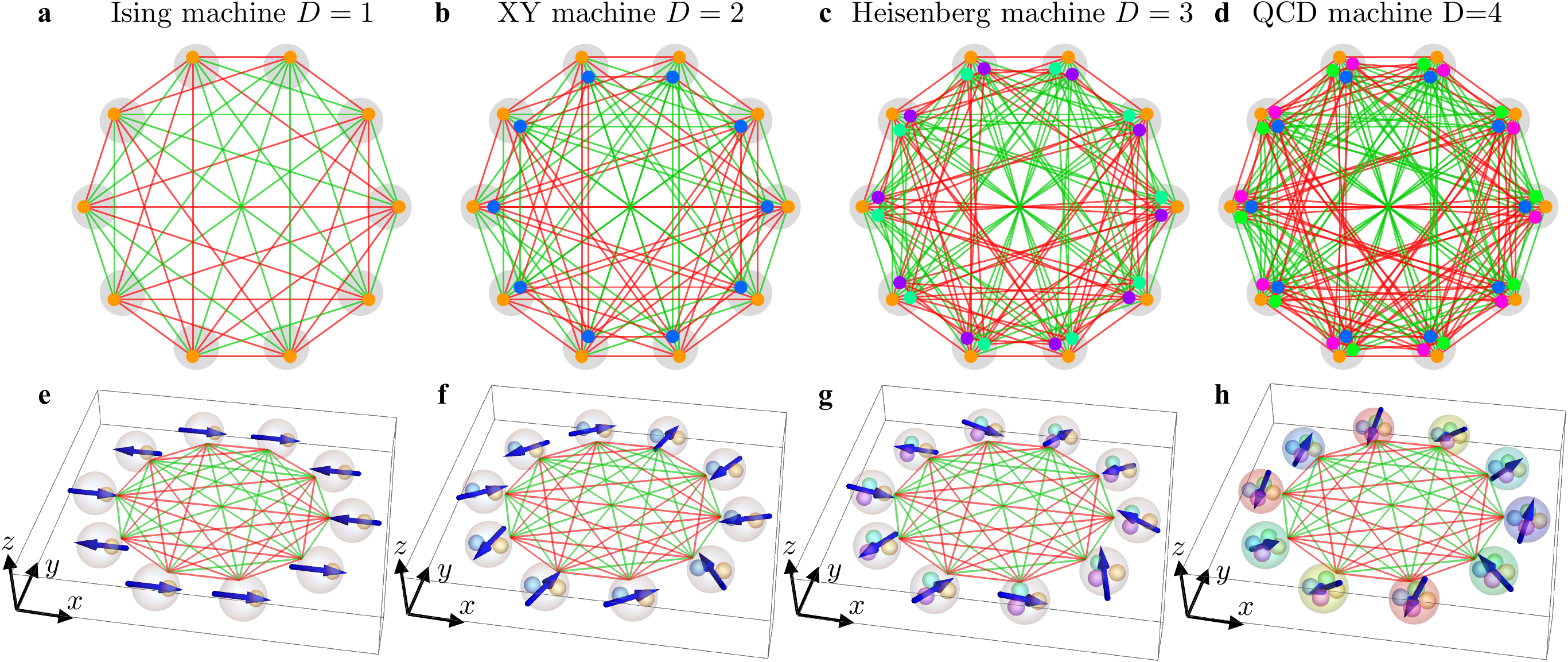}
\caption{Network of $D\times N$ POs simulating a network of $D$-dimensional hyperspins coupled as a random complete K graph. The network is shown with $N=10$ and embedded in a circular geometry. \textbf{a,b,c,d} Full PO network for $D=1,2,3,4$ as in the legends. Green and red lines represent positive end negative entries of the adjacency matrix $\mathbf{J}$, respectively. \textbf{e,f,g,h} Hyperspin representation in the $xyz$-space of the PO network, where hyperspins are represented as in Fig.~\ref{fig:opomultiplet1}. The steady-state values of the real part of the PO amplitudes dynamics $X_j(t)$ from the numerical integration of Eq.~\eqref{eq:equationofmotionmultipletsaa7} determine the state of the spins (details and movies of the dynamics are given in the SI).}
\label{fig:opomultiplet2}
\end{figure*}

\begin{figure*}
\centering
\includegraphics[width=17.9cm]{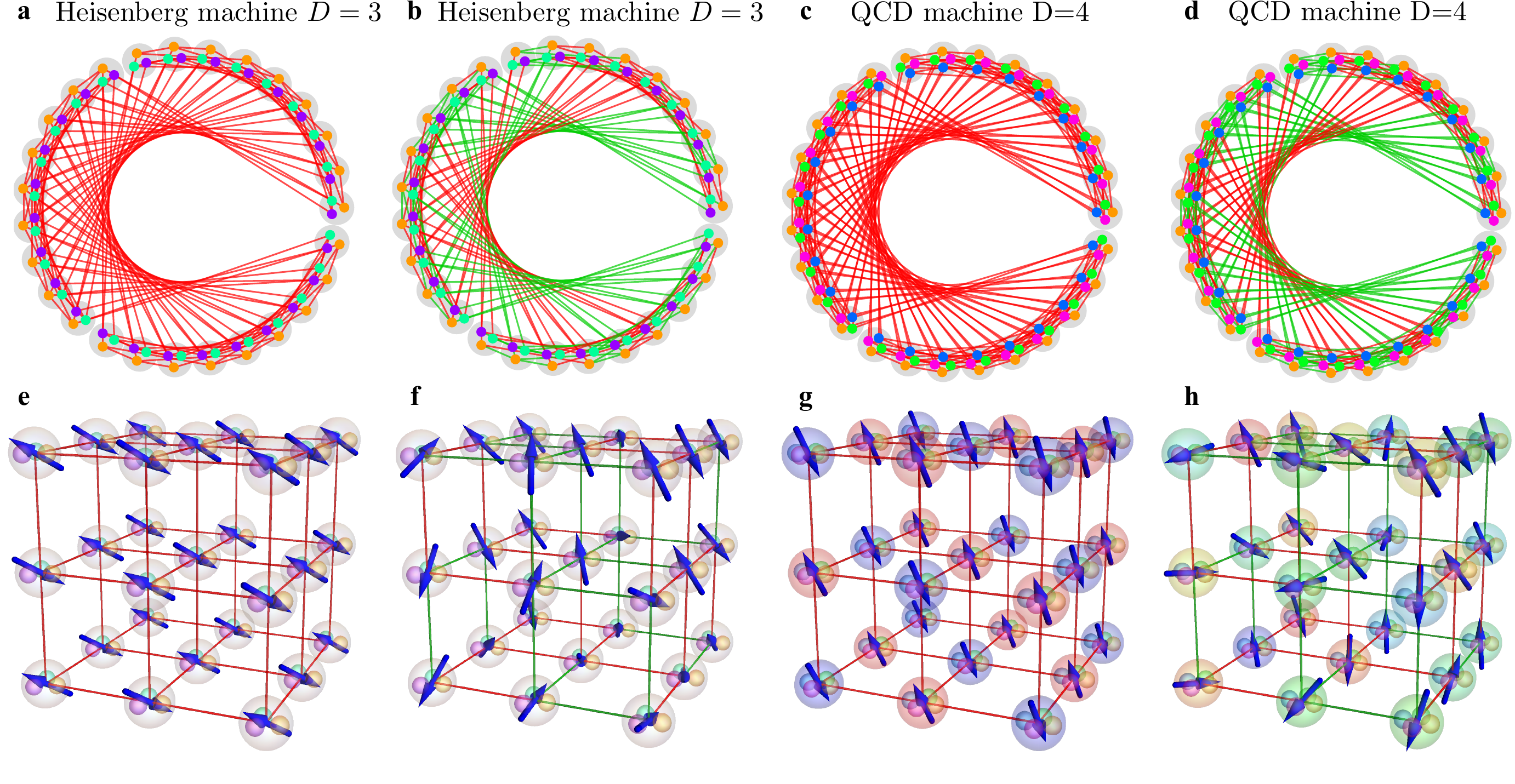}
\caption{Network of $D\times N$ composite POs representing a three-dimensional solid topology with nearest-neighbour interaction (spin glass) of $N=N_x\times N_y\times N_z=27$ spins, here specifically with $N_x=N_y=N_z=3$, and $D=3,4$ as in the legends. \textbf{a,b,c,d}, PO network connectivity in circular embedding for \textbf{a,c} uniform antiferromagnetic interaction, and \textbf{b,d} random binary interaction. \textbf{e,f,g,h}, Hyperspin network representation with solid embedding of the PO connectivity in panels \textbf{a,b,c,d}. The final hyperspin state is retrieved as in Fig.~\ref{fig:opomultiplet2}.}
\label{fig:opomultiplet3}
\end{figure*}

Left panels in Fig.~\ref{fig:opomultiplet1}\textbf{g,h,i,j} show the composite PO as in panel \textbf{b}, and right panels give a three-dimensional representation of the spin $\vec\sigma$ in Table~\ref{tab:oposandspindimension1} in standard hyperspherical coordinates with unit radius~\cite{kalnins2002ndimensional,jingjing2011ndimensional}: The sign of the PO quadrature for $D=1$, and polar and spherical coordinates of the quadrature unit vector for $D=2$ and $D=3$ respectively. For $D=4$, the arrow represents the ``meson'' component in three-dimensional spherical coordinates, while the additional angle encoding the ``scalar'' coordinate is a color assigned to the outer sphere.

\vspace{0.4cm}
\noindent
\textbf{Coupled $D$-dimensional hyperspins}\\
We now move to the case of coupled composite POs and explicit the relation between the network dynamics and the $D$-vector spin model Hamiltonian~\cite{RevModPhys.71.S358}
\begin{equation}
H_D(\{\vec{\sigma}\})=-\sum_{q,p=1}^{N}J_{ql}\,\vec{\sigma}_q\cdot\vec{\sigma}_p \,\, ,
\label{eq:equationofmotionmultipletsaa17}
\end{equation}
with non-uniform hyperspin-hyperspin coupling quantified by the adjacency matrix $\mathbf{J}$. The system of $N$ coupled POs is modeled by the $D\times N$ classical equation of motion
\begin{equation}
\ddot x_j\!+\!\left[1\!+\!gh\!\left(\!1\!-\!\beta\sum_{l=1}^{DN}W_{jl}x^2_l\!\right)\!\sin(2t)\right]\!x_j\!-\!g\sum_{l=1}^{DN}C_{jl}\dot{x}_l\!=\!0\,,
\label{eq:equationofmotionmultipletscoupled1}
\end{equation}
where $C_{jj}=-1$ identifies the intrinsic loss, and the matrix $\mathbf{W}$ organizes the POs as $N$ hyperspins of $D$ commonly pumped POs. In this arrangement, the $q$-th hyperspin is identified by the PO indexes $j\in\mathbb{S}_q$ with $\mathbb{S}_q\coloneqq\{1+(q-1)D,\ldots,qD\}$, where each PO amplitude $X_j$ within this set identifies the $\mu$-th component of the $q$-th hyperspin vector $\vec{S}_q$ as $X_j\rightarrow X_{\mu+(q-1)D}\equiv X^{(q)}_\mu$ (see Fig.~\ref{fig:opomultiplet2a} for a pictorial representation with $N=D=2$). Furthermore, $\mathbf{C}$ denotes the coupling matrix, whose off-diagonal element $C_{jl}$ quantifies the coupling strength between any two POs $x_j$ and $x_l$. The coupling matrix can in general be written as the sum of a symmetric and antisymmetric part, identifying the dissipative and energy-preserving part of the coupling, respectively~\cite{PhysRevA.100.023835,Calvanese_Strinati_2020}. Dissipative couplings are commonly considered when using POs for optimiziation~\cite{yamamoto2020isingmachine}, while energy-preserving couplings inducing persistent coherent beats between POs~\cite{PhysRevLett.123.083901} have been recently proposed to realize photonic spiking neurons~\cite{takesue2021spiking}. Hereafter, we focus on symmetric coupling matrices.

The equations for the slowly-varying amplitudes $X_j$ from Eq.~\eqref{eq:equationofmotionmultipletscoupled1} are detailed in the SI. Figure~\ref{fig:opomultiplet2a} shows the arrangement as $N$ hyperspins with $D$ POs. We decompose the coupling matrix as $\mathbf{C}=\mathbf{J}\otimes\mathbf{G}$, where $\mathbf{J}$ is the $N\times N$ adjacency matrix encoding the specific multidimensional spin model, and $\mathbf{G}$ is a $D\times D$ metric tensor. With this choice of $\mathbf{C}$ and redefinition of the indexes, when the dynamics of the PO amplitudes suppresses their imaginary parts, one can write ($j\in\mathbb{S}_q$)
\begin{equation}
\frac{dX^{(q)}_\mu}{dt}\!=\!\left(\!\frac{h}{4}\!-\!\frac{1}{2}\!-\!\frac{h\beta}{2}\!S^2_q\right)\!\!X^{(q)}_\mu+\frac{1}{2}\sum_{p=1}^{N}\sum_{\nu=1}^{D}\!J_{qp}\,G_{\mu\nu}X^{(p)}_\nu,
\label{eq:equationofmotionmultipletsaa7}
\end{equation}
where $S^2_q=\sum_{l\in\mathbb{S}_q}X^2_l$ denotes the amplitude of the $q$-th spin vector. For a given adjacency matrix $\mathbf{J}$ and in the proper regime of pump amplitude $h$ above $h_{\rm th}$, the PO network in Eq.~\eqref{eq:equationofmotionmultipletsaa7} behaves as a gradient descendent system driving the spin configuration towards the minimum of the $D$-vector spin model Hamiltonian in Eq.~\eqref{eq:equationofmotionmultipletsaa17} when $\mathbf{G}=\mathbb{1}_D$ and with the spin vectors $\vec{\sigma}_q=\vec{S}_q/S_q$ (see SI).

We show in Figs.~\ref{fig:opomultiplet2} and~\ref{fig:opomultiplet3} two prototype examples of PO connectivity and equivalent representation as hyperspins in the $xyz$-space. In Fig.~\ref{fig:opomultiplet2}, we consider a random complete (K) graph~\cite{schneider1993graphs} with $N=10$ spins for dimension $D=1,2,3,4$. The adjacency matrix has entries with fixed amplitude $|J_{qp}|=0.03$ and sign randomly chosen with equal probability for each $q$ and $p$. The case $D=1$ in panel \textbf{a} represents the Ising model, and Eq.~\eqref{eq:equationofmotionmultipletsaa7} gives the PO dynamics of CIMs~\cite{s41534-017-0048-9}. In this case, each spin takes a binary value, represented by an oriented arrow along the $x$-axis in panel \textbf{e}. The spin state is retrieved from the steady-state values of the PO amplitudes $\overline{X}_j\equiv\overline{X}^{(q)}_\mu$ from the numerical integration of the complex amplitude equations, whose real-part evolution is Eq.~\eqref{eq:equationofmotionmultipletsaa7}, seeded with a random complex initial condition $X_j(0)$ (details and movies of the hyperspin dynamics are shown in the SI). The higher-dimensional cases in panels \textbf{b,c,d} for $D=2,3,4$ simulate the XY, Heisenberg, and QCD model, respectively. The PO connectivity $\mathbf{C}=\mathbf{J}\otimes\mathbb{1}_D$ for the scalar product in Eq.~\eqref{eq:equationofmotionmultipletsaa17} is obtained by connecting a dot of a given color within a multiplet (gray circle) to the dot of the same color in another multiplet. The spin states in panels \textbf{f,g,h} are the representation of the spin vectors in standard hyperspherical coordinates~\cite{kalnins2002ndimensional,jingjing2011ndimensional} (see Fig.~\ref{fig:opomultiplet1}). In all these cases, the random orientation of the spins reflects the disordered nature of the graph.

In Fig.~\ref{fig:opomultiplet3}, we show a hyperspin glass~\cite{parisi1987spinglass}, i.e., a solid three-dimensional system of $N$ spins in the $xyz$-space in dimension $D=3$ and $D=4$ with nearest-neighbour interaction, arranged as a lattice of $N=N_x\times N_y\times N_z$ hyperspins. Panels \textbf{a,e} and \textbf{c,g} consider a uniform antiferromagnetic interaction, while panels \textbf{b,f} and \textbf{d,h} are with a random binary interaction, where as for the K graphs $|J_{qp}|$ is fixed and its sign is randomly chosen with equal probability. The spin state is obtained as in Fig.~\ref{fig:opomultiplet2}. For the antiferromagnetic interaction, we obtain from our simulations an antiferromagnetically oriented spin structure, represented by the arrows both for $D=3$ and $D=4$ with additional alternating sphere colors. For the other cases (panels \textbf{b,f} and \textbf{d,h}), as for the K graph in Fig.~\ref{fig:opomultiplet2}, the spin orientation is random due to the disordered interaction. It is important to remark that general spin models have impact in many fields. Notable examples include the Ising~\cite{PhysRevResearch.2.043241} and the Heisenberg spin glass~\cite{Baity_Jesi_2019} for $D=1$ and $D=3$, respectively, and the finite-temperature phase transition in QCD with two light-quark flavors for $D=4$~\cite{PhysRevD.29.338,PELISSETTO2002549,PhysRevD.85.094506,PhysRevLett.123.062002}.

\begin{figure}
\centering
\includegraphics[width=8.6cm]{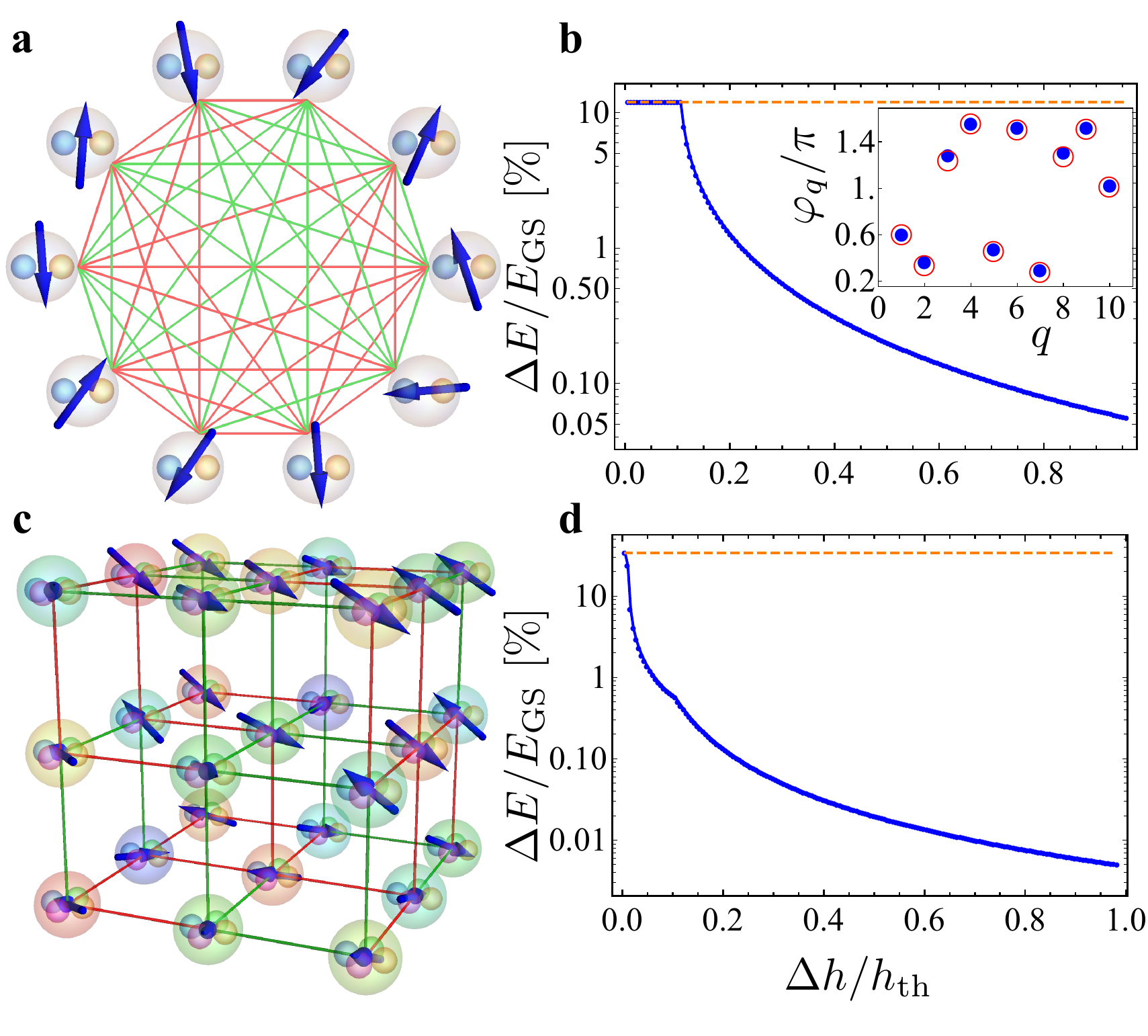}
\caption{$D$-vector Hamiltonian minimization for \textbf{a,b} the XY model ($D=2$) with $N=10$ and complete random K graph in Fig.~\ref{fig:opomultiplet2}, and \textbf{c,d} the QCD spin-glass model ($D=4$) with $N=27$ in Fig.~\ref{fig:opomultiplet3}. Panels \textbf{b} and \textbf{d} show respectively for panels \textbf{a} and \textbf{c} the spin relative energy difference (in percentage) between the energy computed from the PO amplitudes and the GS energy by numerical minimization of the selected XY and QCD Hamiltonian, as a function of the pump amplitude deviation from threshold. The horizontal orange dashed lines mark the spin energy from the eigenvector of $\mathbf{C}$ with largest eigenvalue. Blue dots and red open circles in the inset of panel \textbf{b} are the XY phases $\varphi_q$ from the POs for $\Delta h/h_{\rm th}=0.8$, and the GS phases by numerically minimizing the XY Hamiltonian, respectively. By increasing the pump amplitude, the energy from the POs rapidly approaches the computed GS energy, with determined deviation below $0.1\%$ for the case in panel \textbf{b}, and below $0.01\%$ for that in panel \textbf{d}.}
\label{fig:opomultiplet4}
\end{figure}

\vspace{0.4cm}
\noindent
\textbf{Hyperspin Hamiltonian minimization}\\
We now explicit the working principle of the hyperspin network simulator. We study specifically the $D$-vector model in Eq.~\eqref{eq:equationofmotionmultipletsaa17}. The PO network dynamics in Eq.~\eqref{eq:equationofmotionmultipletsaa7} for general $D$ drives the system close to the ground-state of the $D$-vector Hamiltonian, sharing similarities with the conventional Ising simulators for $D=1$, but with important differences. The hyperspin structure of $D$ multiplet POs is given by the nonlinear coupling due to common pump saturation. For a pump amplitude $h$ slightly above the threshold, nonlinearities affect the dynamics on a time scale much slower than the rate of energy exchange due to the linear coupling~\cite{hamerlyfristratedchain2016,PhysRevLett.126.143901}. The PO amplitudes $X_j$ freeze to the configuration dictated the eigenvector of $\mathbf{C}$ with largest eigenvalue. Depending on the specific form of $\mathbf{C}$, this configuration may coincide with the one minimizing the cost function. This means that the PO network deterministically solves the selected optimization problem when driven above the threshold. Such a phenomenology allows to conclude that the optimization problem belongs to the polynomial (P) class of computational complexity~\cite{PhysRevApplied.16.054022,kalinin2020}, because finding the ground state of the $D$-vector Hamiltonian reduces to finding the eigenvector of $\mathbf{C}$ with maximal eigenvalue. This is indeed the case of panels \textbf{a,e} and \textbf{c,g} in Fig.~\ref{fig:opomultiplet3} with uniform antiferromagnetic interaction. On the contrary, for NP problems, the pump amplitude has to be increased to let the system explore a larger configuration space. When the spin variables are discrete ($D=1$), this results into finding the correct solution of the Ising model with finite success probability~\cite{1805.05217}.

\begin{figure*}
\centering
\includegraphics[width=17.7cm]{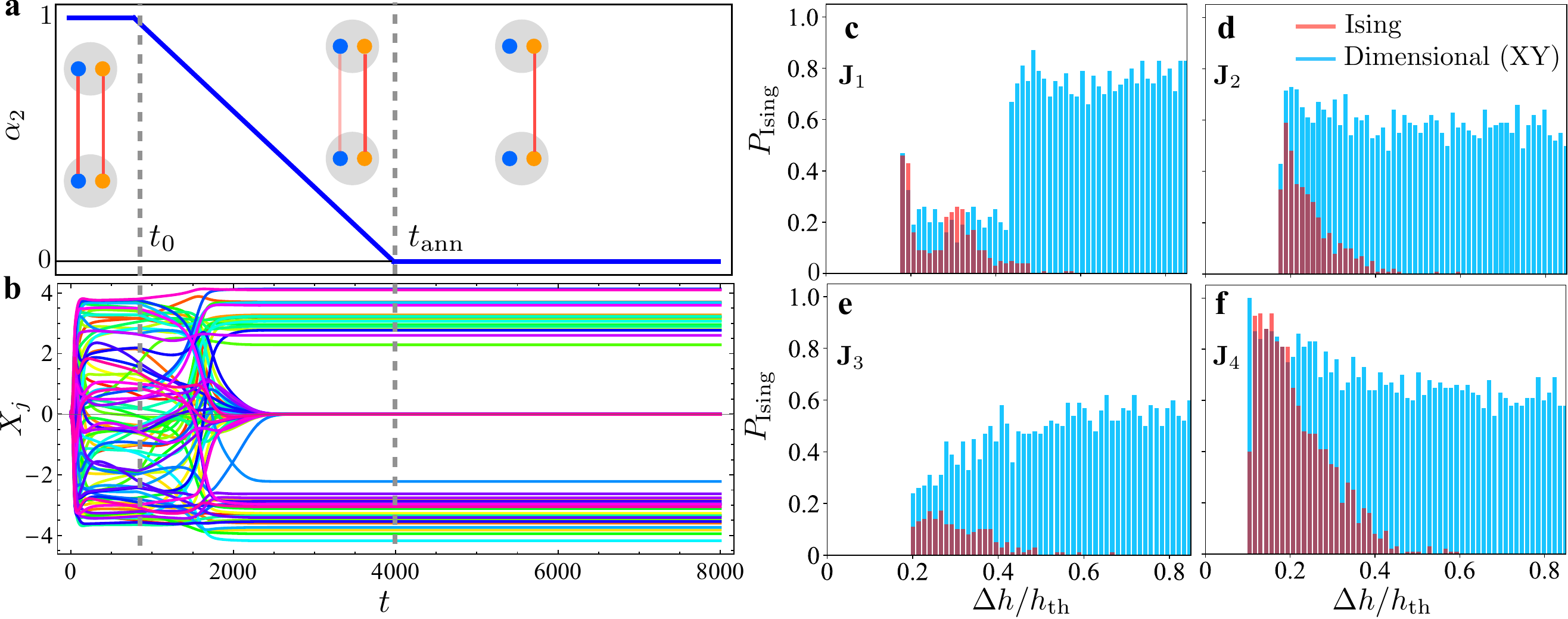}
\caption{Performance comparison between dimensional annealing and discrete spin simulation. We use $N=40$ spins and four random K graphs with adjacency matrices $\mathbf{J}_u$ with $u=1,\ldots,4$. \textbf{a}, Variation of $\alpha_2(t)$ during the annealing. The vertical dashed gray lines mark the starting $t_0$ and final annealing time $t_{\rm ann}$. The insets depict the connectivity between any two XY spins (with negative coupling for illustration purposes). Here, $\alpha_2$ is the metric component along the $y$-axis, which is turned to zero at the end of the annealing protocol. \textbf{b}, Example of dynamics of the real PO amplitudes $X_j$ from Eq.~\eqref{eq:equationofmotionmultipletsaa7} with $\beta=10^{-2}$ during the annealing. After an initial dynamics, the POs corresponding to the $y$-axis components of the XY spins switch off. \textbf{c}-\textbf{f}, Histograms of Ising success probability $P_{\rm Ising}$ from the discrete spin simulation ($D=1$, red histograms), and from the simulation of the XY model with annealing (interpolating between $D=2$ and $D=1$, blue histograms), as a function of the pump power deviation from the threshold value $\Delta h/h_{\rm th}$ and for a given $\mathbf{J}_u$ as in the labels. As evident, the dimensional annealing significantly increases $P_{\rm Ising}$ for a sufficiently large pump amplitude.}
\label{fig:opomultiplet5}
\end{figure*}

For the multidimensional hyperspin case $D\geq2$, the way the PO network performs the optimization of Eq.~\eqref{eq:equationofmotionmultipletsaa17} is shown in Fig.~\ref{fig:opomultiplet4}. We focus specifically on the XY model with K graph in Fig.~\ref{fig:opomultiplet2}\textbf{b,f}, and the random spin glass in Fig.~\ref{fig:opomultiplet3}\textbf{d,h} with $D=4$. The phenomenology is common to other choices of $\mathbf{J}$ (see SI). We show in panels \textbf{b,d} (blue color) the PO energy difference from the computed ground state $\Delta E=E_{\rm PO}-E_{\rm GS}$ as a function of the pump amplitude deviation from threshold $\Delta h=h-h_{\rm th}$. The pump amplitude $h$ varies from the analytical threshold to a numerically-determined value, above which the PO amplitudes acquire a nonzero imaginary part~\cite{PhysRevA.100.023835}. The PO energy $E_{\rm PO}$ is found from Eq.~\eqref{eq:equationofmotionmultipletsaa17} by determining the hyperspins from the PO steady-state amplitudes $\vec{\sigma}_q=(\overline{X}^{(q)}_1,\ldots,\overline{X}^{(q)}_{D})/S_q$, and the ground-state value $E_{\rm GS}$ is found by numerically minimizing Eq.~\eqref{eq:equationofmotionmultipletsaa17} with respect to the real variables $\{X^{(q)}_\mu\}$ using the minimizer \texttt{NMinimize} in Wolfram Mathematica. The horizontal orange dashed line marks the PO energy of the eigenvector of the coupling matrix with largest eigenvalue. We find that the PO energy deviation from the computed ground-state value starts correctly from the eigenvector value and monotonically decreases as the pump amplitude is increased above threshold. For the XY model in panels \textbf{a,b}, we find that the the energy deviation reaches values that are below approximately $0.1\%$, while for the spin-glass QCD model in panels \textbf{c,d}, the energy deviation goes even below approximately $0.01\%$. The inset in panel \textbf{b} shows the phases $\varphi_q/\pi$ of the $N=10$ XY spins, computed from the PO phases (blue filled circles) and from the numerical minimization of the XY Hamiltonian (open red circles), for a pump amplitude $\Delta h/h_{\rm th}=0.8$. As evident, the two data series are overlapped. In light of these results, we conclude that our PO network in Eq.~\eqref{eq:equationofmotionmultipletsaa7} finds to a very good approximation the ground-state of the $D$-vector hyperspin model.

\vspace{0.4cm}
\noindent
\textbf{Dimensional annealing}\\
A notable advantage of the hyperspin machine compared to state-of-the-art continuous spin simulators is the ability to define a spin according to its Cartesian projections. This opens the possibility to simulate quantum spin models and emulate several quantum-inspired algorithms to solve optimization problems using a purely classical system. We now discuss one of such remarkable applications, i.e., solving the Ising model by performing an annealing protocol starting from the XY model. This application follows from using a time-dependent diagonal metric tensor $\mathbf{G}(t)={\rm diag}(\alpha_1,\ldots,\alpha_D)$, where $\alpha_\mu=\alpha_\mu(t)$ is a time-dependent metric component. Starting from the XY Hamiltonian [Eq.~\eqref{eq:equationofmotionmultipletsaa17} with $D=2$] at time $t=0$, which is for $\mathbf{G}={\rm diag}(1,1)$, we arrive at the Ising Hamiltonian [Eq.~\eqref{eq:equationofmotionmultipletsaa17} with $D=1$] for times $t$ larger than a given ``annealing'' time $t_{\rm ann}$, above whith $\mathbf{G}={\rm diag}(1,0)$. We use $\alpha_1=1$ and independent of $t$. For the other time-dependent metric component, we take $\alpha_2(t)=1$ for a time $t$ smaller than a fixed $t_0$, which is the starting time of the annealing procedure, and $\alpha_2(t)=0$ for $t>t_{\rm ann}$. For an intermediate time between $t_0$ and $t_{\rm ann}$, the metric component linearly interpolates between $1$ and $0$ (see Fig.~\ref{fig:opomultiplet5}\textbf{a}). In this way, the PO network simulates $H_{\rm XY}$ for a time $t<t_0$, it reduces to $H_{\rm Ising}$ for $t>t_{\rm ann}$, while for $t$ between $t_0$ and $t_{\rm ann}$, it interpolates between the two models, i.e., $H(s)=(1-s)H_{\rm XY}+sH_{\rm Ising}$, where $s=(t-t_0)/(t_{\rm ann}-t_0)$. The resulting PO amplitude dynamics is shown in panel \textbf{b}. During an initial nontrivial dynamics for a time smaller than $t_0$, the PO network simulates the XY model starting from random initial conditions. In this first stage, as shown in Fig.~\ref{fig:opomultiplet4}, the system starts to converge towards the minimum of the XY Hamiltonian. For a time larger than $t_0$, the reduction of the $\alpha_2$ metric causes the POs corresponding to the $\mu=2$ components (i.e., $y$) to gradually switch off. The other PO amplitudes defining the $\mu=1$ components (i.e., $x$) converge to a steady state following a dynamics dominated by the Ising Hamiltonian. After the annealing procedure, the XY spins are polarized along the $x$-axis and represent an Ising state (see Fig.~\ref{fig:opomultiplet2}). The reason why we start the annealing procedure after a finite time $t_0$ is to let the PO amplitudes be amplified sufficiently above the initial random values before reducing the system dimensionality. We remark that the annealing protocol proposed here differs from conventional quantum annealing, where one seeks for the ground state of the classical Ising model with $z$-aligned spins starting from a configuration along an orthogonal direction ($x$ or $y$)~\cite{PhysRevE.58.5355}. Our protocol performs a ``dimensional crossover'' between two $D$-vector models with the same adjacency matrix $\mathbf{J}$ but in different dimension, specifically from $D=2$ to $D=1$. As such, we name our protocol as dimensional annealing.

We now focus on a network of $N=40$ spins and show that the dimensional annealing dramatically increases the success probability to solve the Ising model. To reach this goal, we proceed as follows. We choose four adjacency matrices $\mathbf{J}_u$ with $u=1,2,3,4$ representing four random complete K graphs with binary edge weights $|J_{qp}|=0.02$. For each adjacency matrix, we repeat the numerical integration of the PO amplitudes equations for $D=1$ a number $M=100$ of times, and retrieve for each run the Ising spin values from the steady-state amplitudes as described before. From the obtained phases, the $M$ Ising energies $E_{m,{\rm PO}}^{\rm(Ising)}$ are computed, where $m=1,\ldots,M$. The success probability $P_{\rm Ising}$ is defined as the number of runs such that $E_{m,{\rm PO}}^{\rm(Ising)}=E^{\rm(Ising)}_{\rm GS}$, divided by $M$. To find the global Ising ground-state energy $E^{\rm(Ising)}_{\rm GS}$, we resort to a Monte-Carlo Metropolis-annealing inspired algorithm~\cite{metropolis1953}. We remark that, differently from the cases $D\geq2$ in Fig.~\ref{fig:opomultiplet4}, we cannot here resort to the numerical minimization of $H_{\rm Ising}$ because the minimization is more likely to get stuck in local minima due to discrete nature of the spin variables for $D=1$. The computation of $P_{\rm Ising}$ is performed for different values of the pump amplitude deviation from threshold $\Delta h/h_{\rm th}$ and plotted as red histograms in panels \textbf{c,d,e,f} of Fig.~\ref{fig:opomultiplet5}. As evident, the success probability is nonzero only in a narrow range of $h>h_{\rm th}$, and the details of the histograms critically depend on the coupling matrix. These observations are consistent with those in Ref.~\cite{PhysRevLett.126.143901}. The fact that the PO network does not find the global solution of the Ising Hamiltonian is a signature of the NP-hard nature of the optimization problem. As the pump amplitude increases, the success probability decreases. This fact is ascribed to the heterogeneity of the amplitudes~\cite{PhysRevLett.122.040607}: The PO system explores a larger configurational space, and the probability to converge to the global minimum of the Ising model decreases.

\begin{figure}[t]
\centering
\includegraphics[width=8cm]{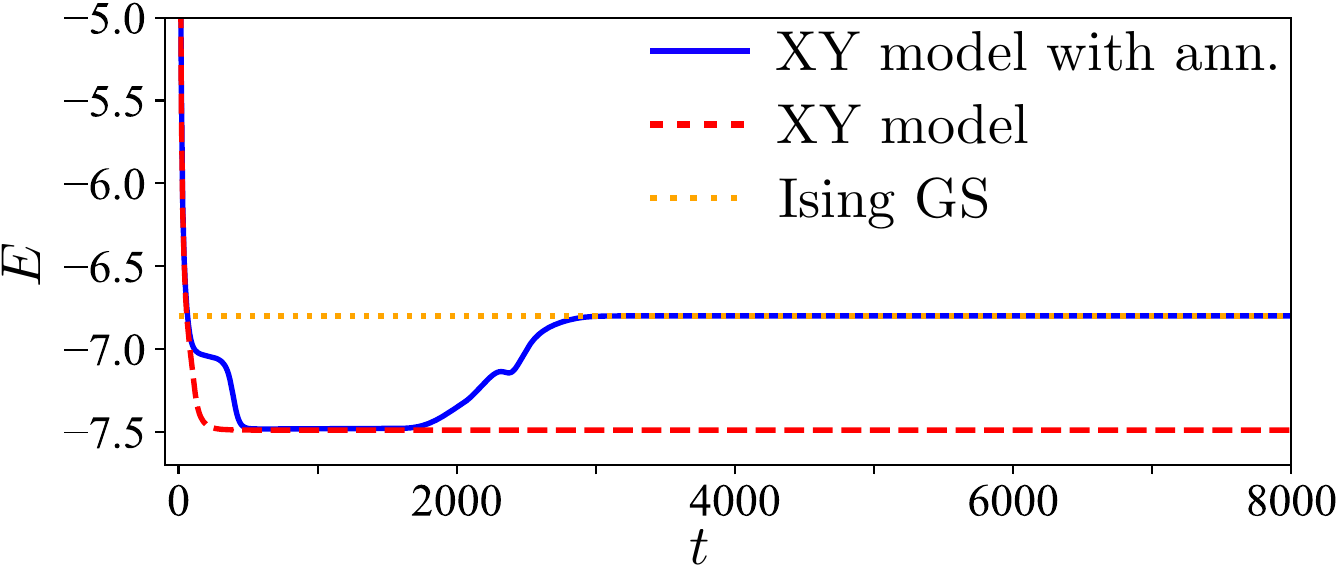}
\caption{Dynamics of the spin energy $E(t)$ from the XY model simulation ($D=2$, red dashed line) and the XY model with dimensional annealing (from $D=2$ to $D=1$, blue line). The horizontal dotted line marks the calculated ground-state energy of the Ising model ($D=1$). The energy first tends to reach the steady state of the XY model. Subsequently, the dimensional annealing drives the energy to a new steady state of the Ising Hamiltonian, which is the Ising ground state, found at a higher energy compared to the XY steady-state energy.}
\label{fig:opomultiplet6}
\end{figure}

We then simulate the XY model with dimensional annealing for the same adjacency matrices $\mathbf{J}_u$, and compute the success probability of the Ising model (see panel \textbf{a,b}). The resulting histograms are shown in blue in panels \textbf{c,d,e,f}, and compared to the red histograms computed for $D=1$. The success probability for the dimensional annealing (blue) stays above $50\%$ even for large pump amplitudes, where the corresponding value from the discrete spin simulation (red) is negligible. This remarkable result is a consequence of the fact that the hyperspin machine finds the state of a \emph{discrete} spin model from the dynamics of a \emph{continuous} spin system that \emph{gradually} reaches in time the target discrete model Hamiltonian. This has a twofold advantage in terms of increasing the probability to find the global minimum of the discrete model: First, the Ising ground-state configuration (i.e., the state with all spins oriented along the same direction) is a particular excited state of the larger class of XY states (i.e., spins taking any orientation on the $xy$-plane). This fact is exemplified in Fig.~\ref{fig:opomultiplet6}, where the time variation of the energy $E$ from the XY model with and without dimensional annealing is shown. After a first dynamical transient where the energy tends to the steady-state value of the XY model, the dimensionality reduction drives the energy to a minimum of the Ising Hamiltonian, at higher energy compared to the XY model steady-state value. As such, local minima of the energy landscape can be smoothly escaped by exploiting the additional dimension starting from an energy value that is in general below the target one. In contrast, escaping a local minimum in the discrete model itself is harder since it can occur only by full spin flips, which intrinsically requires to overcome a larger stiffness compared to the continuous case. Second, the additional local minima introduced by the second dimension are gradually eliminated in time by the dimensional crossover. Therefore, the final minimum found by the annealing is by construction a minimum of the Ising Hamiltonian.

\vspace{0.6cm}
\noindent
\textbf{Summary and perspectives}\\
We propose and theoretically validate a network of coupled POs to simulate systems of hyperspins in general dimension $D$. An isolated hyperspin is realized by feeding $D$ POs with the same pump field, forming a PO multiplet, and a network of coupled hyperspins is achieved by coupling POs belonging to different multiplets. Focusing on PO connectivities implementing the standard Euclidian scalar product, we show that our system converges close to the minimum of the $D$-vector spin Hamiltonian. An advantage of our proposal is that we construct an hyperspin from its Cartesian coordinates, each represented by a specific PO in the multiplet. Thus, we can implement spin models with arbitrary connectivity and emulate quantum algorithms on a purely classical system. We exploit this feature to propose a dimensional annealing protocol, which interpolates between the XY and Ising Hamiltonians. We show that our protocol significantly enhances the success probability to find the global minimum of the Ising Hamiltonian for selected coupling matrices. Intriguing future developments will be the implementation of effective magnetic fields, whose realization with POs for the Ising model has been proposed in Ref.~\cite{PhysRevApplied.13.054059}, as well as the simulation of nonzero temperature in a controllable way~\cite{Takeda_2017}. The hyperspin machine paves the way towards the numerical and experimental study of previously unaccessible critical phenomena in advanced spin models, as well as the simulation of quantum spin models like the Ising model in a transverse field~\cite{PFEUTY197079} at an unprecedented scale. In this manuscript, we focus on the $D$-vector spin model, but our system allows the implementation of general spin Hamiltonians where a PO of a given spin is connected to any other PO in another spin, i.e.,
\begin{eqnarray}
H_{\rm spin}(\{\vec{\sigma}\})=-\sum_{q,p=1}^{N}J_{qp}\sum_{\mu,\nu=1}^{D}G_{\mu\nu}\,\sigma^{(q)}_\mu\sigma^{(p)}_\nu \,\, ,\\\nonumber
\label{eq:spinhamiltoniangeneraldimension1}
\end{eqnarray}

\vspace{-0.2cm}
The hyperspin machine can hence simulate spin models with anisotropic interactions. A relevant case is with $D=3$, which describes the anisotropic Heisenberg model with symmetric and Dzyaloshinsky-Moriya interactions stabilizing nontrivial magnetic textures in solids~\cite{PhysRevB.103.174422,DZYALOSHINSKY1958241,PhysRev.120.91}. Furthermore, in this manuscript, we focus on identical PO multiplets. However, the hyperspin machine allows multiplets of any size within the same network, opening the possibility to realize models with hybrid symmetries~\cite{mathey2013xyising}. The design of the hyperspin machine with POs opens the future perspective to experimentally realize fully-optical, scalable, and size-independent continuous spin simulators, extending recent proposals with an optical cavity with a nonlinear medium and spatial light modulators, similar to that in Ref.~\cite{PhysRevApplied.16.054022} for the Ising model.

\vspace{0.4cm}
\noindent
\textbf{Acknowledgements}\\
We thank Davide Pierangeli for fruitful discussions.


%

\end{document}